\documentclass[journal]{vgtc}                     % final (journal style)
\onlineid{1000}

\vgtccategory{Research}

\vgtcpapertype{system}

\title{Interactive Public Transport Infrastructure Analysis through Mobility Profiles: Making the Mobility Transition Transparent}

\author{%
  \authororcid{Yannick~Metz}{0000-0001-5955-4487},
  {Dennis~Ackermann},
  \authororcid{Daniel~A.~Keim}{0000-0001-7966-9740}, and
  \authororcid{Maximilian~T.~Fischer}{0000-0001-8076-1376}
}

\authorfooter{
  \item
  	Y. Metz, D. Ackermann, D. A. Keim, M. T.\ Fischer are with the University of Konstanz. E-mail: \{yannick.metz, dennis-fabian.ackermann, keim, max.fischer\}@uni-konstanz.de.
}

\abstract{%
    Efficient public transport systems are crucial for sustainable urban development as cities face increasing mobility demands.
    Yet, many public transport networks struggle to meet diverse user needs due to historical development, urban constraints, and financial limitations.
    Traditionally, planning of transport network structure is often based on limited surveys, expert opinions, or partial usage statistics.
    This provides an incomplete basis for decision-making.
    We introduce an data-driven approach to public transport planning and optimization, calculating detailed accessibility measures at the individual housing level.
    Our visual analytics workflow combines population-group-based simulations with dynamic infrastructure analysis, utilizing a scenario-based model to simulate daily travel patterns of varied demographic groups, including schoolchildren, students, workers, and pensioners.
    These population groups, each with unique mobility requirements and routines, interact with the transport system under different scenarios traveling to and from Points of Interest (POI), assessed through travel time calculations.
    Results are visualized through heatmaps, density maps, and network overlays, as well as detailed statistics.
    Our system allows us to analyze both the underlying data and simulation results on multiple levels of granularity, delivering both broad insights and granular details.
    Case studies with the city of Konstanz, Germany reveal key areas where public transport does not meet specific needs, confirmed through an initial user study.
    Due to the high cost of changing legacy networks, our analysis facilitates the identification of strategic enhancements, such as optimized schedules or rerouting, and few targeted stop relocations, highlighting consequential variations in accessibility to pinpointing critical service gaps.
    Our research advances urban transport analytics by providing policymakers and citizens with a system that delivers both broad insights with granular detail into public transport services for a data-driven quality assessment at housing-level detail.
}


\teaser{
  \centering
  \includegraphics[width=\linewidth, alt={Our workflow for the creation of mobility profiles for the analysis and improvement of public transport infrastructure}]{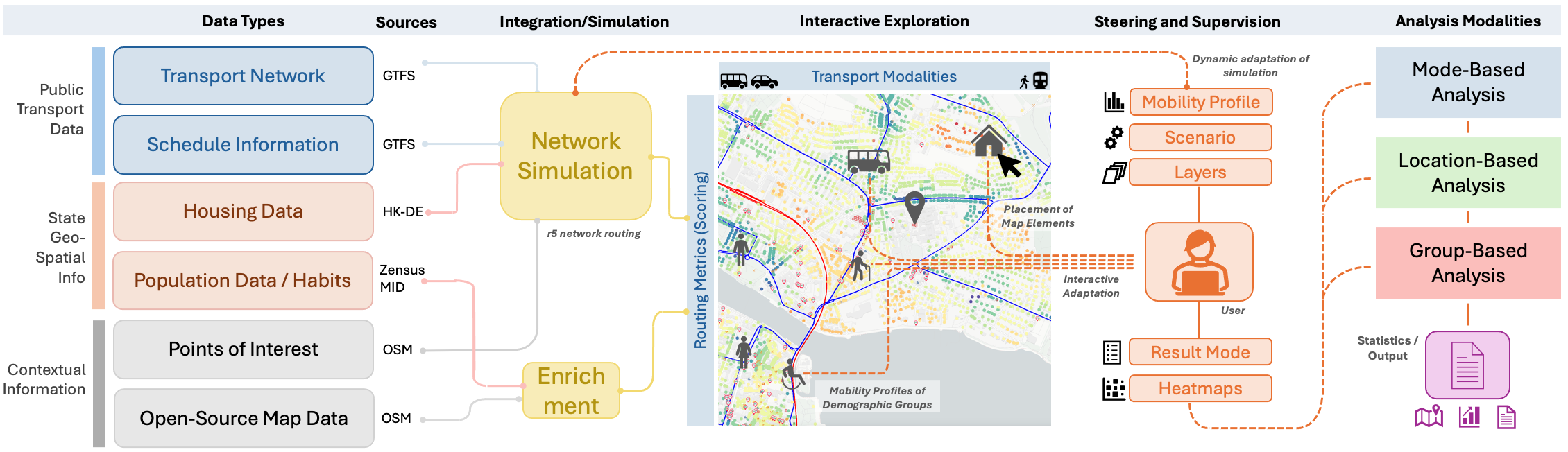}
  \caption{%
  	\textbf{Our workflow for the creation of mobility profiles for the analysis and improvement of public transport infrastructure:}
   The approach combines multiple open data sources and integrates them together with aggregated network flow simulations from existing frameworks~\cite{Pereira.R5.2021} trough an enriched map visualization to assess the quality, i.e. connectedness and travel times, of public transport at housing-level detail.
   Exploration and steering, changing how persons move between their homes and various POIs as part of their daily routine, is enabled by interactively adapting simulation constraints and aggregations.
   Further, it allows us to analyze the effect of network changes on the population's connectivity.
   Our demonstrator enables multiple analysis modalities:
   As output, heatmap visualizations, usage statistics, and quantitative reports can be extracted, detailing which transportation \textit{modes} work, how accessible \textit{locations} are concerning travel- and wait times, and how good the network reflects the need of different population \textit{groups}.
  }
  \label{fig:teaser}
}

\graphicspath{{figs/}{figures/}{pictures/}{images/}{./}} % where to search for the images
\usepackage{mathptmx}                  % use matching math font
\usepackage{enumitem}
\usepackage{amsfonts}
\usepackage{amsmath}
\usepackage{wrapfig}

\renewcommand{\manuscriptnotetxt}{%
\vspace*{-0.1cm}
}

\begin{document}

%%%%%%%%%%%%%%%%%%%%%%%%%%%%%%%%%%%%%%%%%%%%%%%%%%%%%%%%%%%%%%%%
%%%%%%%%%%%%%%%%%%%%%% START OF THE PAPER %%%%%%%%%%%%%%%%%%%%%%
%%%%%%%%%%%%%%%%%%%%%%%%%%%%%%%%%%%%%%%%%%%%%%%%%%%%%%%%%%%%%%%%

%% ``\begin{document}'' command.
% It prepares and prints the title block.
\firstsection{Introduction}
\label{sec:introduction}

\maketitle

Efficient public transport systems are a foundational pillar in urban development to meet both increasing mobility demands by mobile and economically active citizens~\cite{Miller.PublicTransportSustainability.2016} as well as to reduce emissions on the pathway to a green transition and net-zero to achieve global sustainability goals~\cite{Schlacke.ImplemetingEUFitfor55.2022}, as highlighted by the European Union's (EU) Climate Law’s program \emph{Fit for 55}~\cite{EU.FItfor55.2021}. Public transport is pivotal not only in mitigating traffic congestion and reducing environmental footprints but also in reducing economic cost and social inclusiveness~\cite{Ovaere.CostEffectiveEUTransport.2022}.
As urban areas expand and the global population becomes increasingly concentrated in cities, the \textbf{demands on public transportation systems} intensify~\cite{Awad.PostCovidTravelPatternSpain.2021, Gkiotsalitis.PublicTransportPlanningDirections.2021}.
While city design in the last century was predominantly focused on individual mobility with cars, efforts to make cities more livable and decrease travel times through concepts such as the \emph{15-minute city}~\cite{Moreno.15minCities.2021} in Paris, France or changed urban planning like in Amsterdam, The Netherlands~\cite{Feddes.MakingAmsterdamCylcingCity.2020}, have proven to be hugely successful~\cite{Allam.NetZeroUrbanFuture.2022}.
Combined with these transitions, a reduction in individual mobility and increased public transport is often observed.

However, public transport networks often \textbf{struggle to meet} these increasing demands due to constraints imposed by historical development, urban infrastructure, and financial limitations~\cite{Mandl.EvaluationOptimizationPublicTransportation.1980}.
Traditional planning methods for public transport systems typically rely on inherently limited data sources—such as small-scale surveys, expert opinions, and partial usage statistics.
While complex network flow simulations exist~\cite {Axhausen.MATSim.2016, PTV.VISSIM.2024, DLR.SUMO.2024}, traditional urban planning, when based on movement patterns, primarily focuses on optimizing traffic flow for a modeled infrastructure and less on typical usage patterns.
Simultaneously, transport networks are frequently compromised by constraints regarding spatial limitations of dense environments, cost-benefit considerations, and persistent budgetary constraints, making this a complex optimization problem with a vast search space difficult to solve algorithmically.
Due to these difficulties, these methods--while potentially providing some valuable insights--often result in sub-optimal planning and an incomplete understanding of network efficiencies and user needs~\cite{Farahani.ReviewPublicTransportDesignProblems.2013}.

Recognizing these challenges, we propose a \textbf{novel, data-driven framework} (see Figure~\ref{fig:teaser}) that aims to significantly enhance the ability and accuracy of public transport planning by focusing on \emph{travel times for
representative scenarios with household-level granularity}, combining automated analysis with human steering and expertise:
By leveraging existing simulation frameworks for network flow, we introduce the concept of dynamic group \textbf{mobility profiles}, simulating and visualizing the daily travel patterns of diverse demographic groups from their homes to various points of interest (POIs), capturing a detailed picture of urban mobility needs, thereby \textbf{contributing}:

\begin{itemize}[topsep=0pt,itemsep=-.75ex,partopsep=1ex,parsep=1ex]
    \item An \textbf{interactive simulation framework} for assessing and analyzing existing public transport networks
    \item An interactive assessment of \textbf{connectivity and accessibility}, inspectable by scenarios and demographics, at \textbf{housing-level detail}, only relying on open-source data for Points of Interest (POI) and housing (OSM), network, and schedule information (GTFS).
    \item Two \textbf{case studies} for the city of Konstanz, demonstrating the effectiveness in identifying transport deficiencies.
    %\item A discussion on open challenges and research opportunities.
\end{itemize}

This detailed mapping supports the development of targeted interventions like optimized bus routes, strategically placed new stops, and revised transportation schedules through the identification of critical service gaps.
%We designed an interactive application that is both extensive in the provision of analysis capabilities, and easy to use even for non-expert users.
The accessibility of our system opens up public transport planning to \textbf{wider audiences}, strengthening data-based public discourse.
%These proposed enhancements are designed to be both cost-effective and impactful, addressing the most pressing needs without necessitating extensive infrastructural overhauls.
%The contributions of this research extend beyond academic discussion to offer practical tools and actionable insights.
In \textbf{conclusion}, this work bridges the gap between theoretical planning and practical implementation:
By providing a more precise and scalable data-driven methodology for assessing and improving public transport services, we aim to support policymakers and city scientists in making informed decisions.
Further, we set the stage for future research to incorporate multi-modal transport scenarios and expand beyond local to national scales, thereby paving the way for more efficient, inclusive, and sustainable public transport systems.

\begin{figure*}[t]
\centering
\includegraphics[width=\linewidth, alt={The user interface of our approach, enabling the exploration and interactive simulation of Mobility Profiles:
    It consists of two main visual elements: a map view with a legend (a) and a sidebar (b-f) with several control components.
    The background information (b), as well as particular network and POI information, can flexibly be toggled.
    The scenarios reflecting the mobility of a set of demographics can be modified (d), compared to different assumptions (c), and interactively partly selected (e). Further general settings are available (f).}]{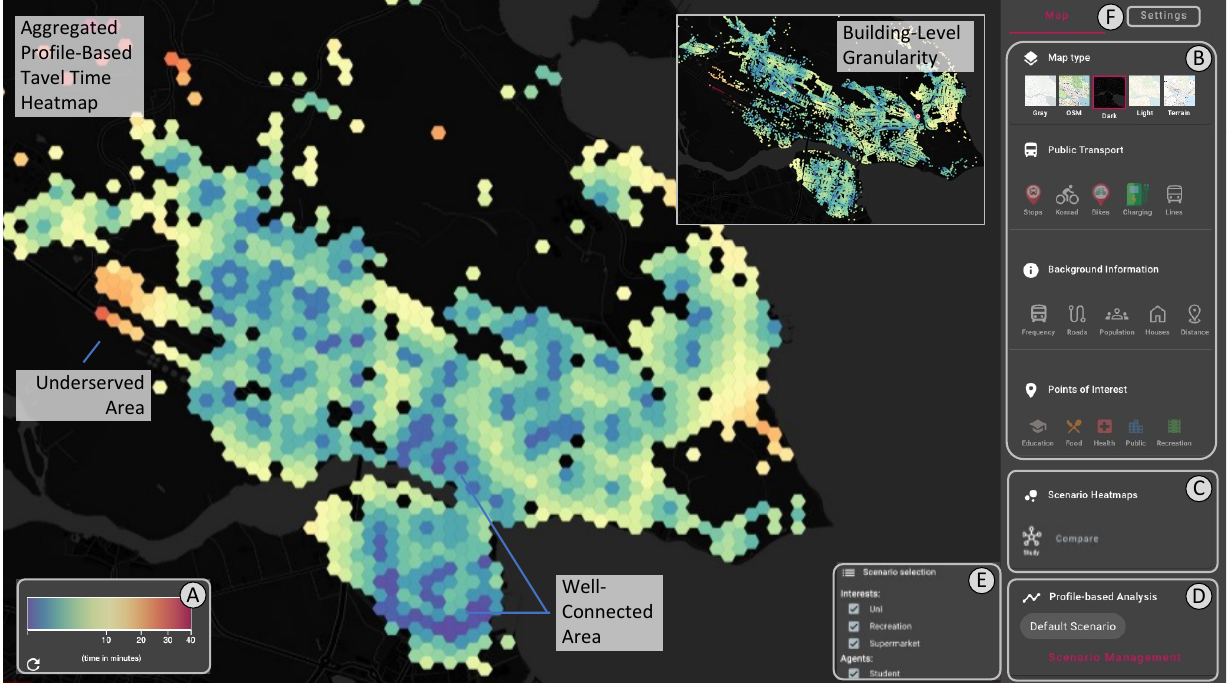}
\caption{The user interface of our approach, enabling the exploration and interactive simulation of Mobility Profiles:
    It consists of two main visual elements: a map view with a legend (a) and a sidebar (b-f) with several control components.
    The background information (b), as well as particular network and POI information, can flexibly be toggled.
    The scenarios reflecting the mobility of a set of demographics can be modified (d), compared to different assumptions (c), and interactively partly selected (e). Further general settings are available (f).}
\label{fig:user_interface}
\end{figure*}

\section{Related Work}
\label{sec:related_work}
Two main research areas are especially relevant to our work: the (interactive) visualizations of public transport infrastructure 
and agent-based simulation for transport optimization.

\subsection{Urban Transport Visualizations}
The creation of public transport infrastructure maps is arguably one of the earliest and most prominent use cases of data visualization, as showcased by abstract representations of subway, train, or bus networks \cite{Avelar2006}.
Much of the academic research on \textbf{Urban Transit Routing Planning (UTRP)} is focused on algorithmic routing and conducted~\cite{Almachi.SurveyVATrafficSimulation.2023} with either the openly available Multi-Agent Transport Simulation (MATSim)~\cite{Axhausen.MATSim.2016}, the commercial Multimodal Traffic Simulation Software (VISSIM)~\cite{PTV.VISSIM.2024} by PTV Group or the open-source Simulation of Urban MObility (SUMO)~\cite{DLR.SUMO.2024} by the German Aerospace Agency (DLR).
These systems primarily focus on the algorithmic routing problem to design networks in their entirety but usually include or integrate with a basic map viewer application to visualize the network structure.
Based on this, there exist multiple commercial applications to analyze public transport infrastructure and passenger data~\cite{pack2010, Jacyna.SimulationPoland.2014, Jacyna.EvaluationTransportAnalyses.2017, Skapinyecz.PossibilitiesTrafficSimulation.2020, Almachi.SurveyVATrafficSimulation.2023}, regarding travel times, incidents, and real-time status of a transport system:
They combine an isochrone map for spatial information with an isotime map flow map for temporal information and a paired journey detail view.
In our work, we go beyond the passive analysis of existing data towards generating novel data via simulation and analyzing changes to the infrastructure interactively.
Burch et al. present \texttt{PasVis}, an interactive application to extend common schematic public transport maps with additional information in the forms of glyphs and extra views showing different statistics such as ridership at certain stations on stops~\cite{Burch2020}.
Zeng et al. use a glyph-based visualization to uncover the relationship between human mobility and point of interests~\cite{Zeng2017}.
We mostly utilize a layer map visualization without extra views and go beyond analyzing existing transport.
Related to our objective of planning and travel time analysis, work by Baltzer et al.~\cite{baltzer2024visualizing} visualizes the effect of new infrastructure on travel time.
In particular, they focus on differences in travel times, represented by isochrone-based maps.
We use related approaches to visualize travel times but move from static pre-computed maps to dynamic and interactive visualizations.

%which, however, lacks visual feedback and explainable features.
The industry leader PTV Group offers several products (e.g., \texttt{PTV Visum} and others) for UTRP, also making a first foray into accessibility simulations with \texttt{PTV Access}~\cite{PTVAccess.2024}, offering grid-level accessibility information.
However, none of the commercial applications are open source, extensible, or freely accessible beyond limited demos.
The recent literature review by Almachi et al.~\cite{Almachi.SurveyVATrafficSimulation.2023} shows the lack of visual analytics systems for such a public transport network assessment.

\subsection{Public Transport Simulation}
UTRP has a long history, with algorithmic approaches for routing~\cite{Mandl.EvaluationOptimizationPublicTransportation.1980}, time tables or schedules~\cite{schiewe2020integrated}.
We are interested in the optimization of infrastructure placement and planning.
Traditional approaches to infrastructure planning have often relied on heuristics, limited data sources, and personal experience.
We advocate for the use of data-driven and interactive planning.
However, because usage and retrospective data are often limited due to privacy concerns and changing existing infrastructure is costly, simulation is a key tool.
Optimization-based approaches have been applied to simple scenarios in the past~\cite{Taber1999}.
To accurately model the behavior of potential passengers and users, agent-based simulations have become popular \cite{crooks2015agent, bastarianto2023agent}.
Incorporating diverse modules and frameworks into agent-based models can more accurately reflect the complex interactions of transportation systems and user behavior~\cite{bastarianto2023agent}.

Calabrò et al. (2020) utilized an ant colony simulation-based optimization to bridge the service gap in weak-demand areas by designing feeder bus routes in Catania, Italy.
This approach leverages agent-based modeling to optimize vehicle routing, demonstrating significant improvements in public transport coverage and ridership by analyzing various scenarios using key performance indicators~\cite{Calabro.WeakDemandAreas.2020}.

Similarily, Manser et al. (2020) extended an agent-based microsimulation framework to design and evaluate large-scale public transport networks for the city of Zurich, Switzerland.
Their study, incorporating dynamic demand responses to network changes, suggests that a redesigned network with smaller vehicles and higher frequencies can increase transit ridership while reducing subsidies. This method also aids in identifying corridors for capacity upgrades~\cite{manser2020designing}.

Yoon et al. (2022) developed an open-source simulation sandbox to assess various transit system designs, from fixed-route services to on-demand microtransit.
By comparing these designs under different demand scenarios, their study highlights the flexibility and adaptability of semi-flexible and on-demand services, enabling for optimizing urban transit systems~\cite{yoon2022simulation}.

Ma and Chow (2022) explored frequency settings in transit networks using a multi-agent simulation to capture activity-based mode switches.
Their work emphasizes the effectiveness of agent-based simulations in fine-tuning transit network frequencies to better match real-world commuter behavior, thus enhancing overall service efficiency~\cite{ma2022transit}.

For a more general reachability analysis, Mocanu et al.
\cite{Mocanu.DataDrivenAnalysisAcessibility.2021} used a simplified synthetic transport model to investigate the mode shift potential from car to public transport in Germany.
Hörl et al. describe a pipeline to generate synthetic travel demand data from open data. They present a synthetic dataset for Paris and its surrounding region~\cite{horl2021synthetic}.

Some work has used mobility profiles and points of interest data in public transport contexts, mainly to enhance existing lines and schedules, but often failing to address underserved neighborhoods or plan new lines~\cite{fabbiani2018analysis}. Li et al. (2018) utilized point of interest data to identify employment locations and commuting patterns, employing a categorization similar to ours, but without directly linking it to public transport planning~\cite{li2018using}. 

In summary, compared to the existing works, we focus on the visualization aspects and the interactive user feedback mechanisms for steering an agent-based simulation.
Our approach uses a standardized data pipeline, leveraging worldwide harmonized data formats, making it modular and easily extendable to other regions in the future.

\section{Methodology}
\label{sec:methodology}
The foundation of our agent-based simulation is geospatial data from diverse but standardized \textbf{open-data sources}, such as public transport network and schedule information, housing, and population data, as well as volunteered geographic information (VGI) data such as points of interest (POIs).
Based on these foundations, we implement a system to interactively explore and adapt transport network simulation data, for which we leverage existing network routing frameworks~\cite{Pereira.R5.2021}.

As primary \textbf{goals}, we aim to (G1)~visualize the quality of existing public transport architecture by geographic location, (G2)~enable a detailed analysis of public transport accessibility for different demographic groups and usage scenarios, and (G3)~simulate the approximate effectiveness of adaptations to the infrastructure.
We demonstrate that we can achieve these goals by only relying on open data sources, enabling our approach to scale.
In the following, we describe details of the data sources, the processing pipeline, and the underlying computations.
A detailed overview of our approach is shown in Figure~\ref{fig:teaser}.

\subsection{Underlying Data and Input}
Public transport analysis can leverage a magnitude of both public and private data and it in general requires transport network and schedule information, infrastructure maps (e.g., streets), as well as flow information.
We utilize a series of well-known public open-data sources for our analysis, relying on GTFS, OSM, as well infering flow based on typical interest and population densities:

\paragraph{Data Sources}
The basis for route planning is formed by country-wide \textit{target timetable data}.
Such timetable data is usually available from local transport operators, and public authorities in a region or country often publish a compiled version in various formats, e.g., the General Transit Feed Specification (GTFS)~\cite{GTFS.2024}.
For Germany, we use a current \textit{GTFS-NeTEx dataset}~\cite{DELFI.2024} from the DELFI integration platform (DIP) available via an open API, providing network, \textbf{timetables}, \textbf{routes}, and \textbf{stop} information.
It contains c. 645,000 stops and 1,600,000 daily trips from 434 transport agencies for various transport modes, including bus, subway, tram, regional and high-speed rail, and ferry.

\begin{figure*}[tbh]
    \centering
    \includegraphics[width=\linewidth, alt={Data Exploration Mode: Map layers to inspect underlying information like transport lines, stop locations, points of interest, and housing data.}]{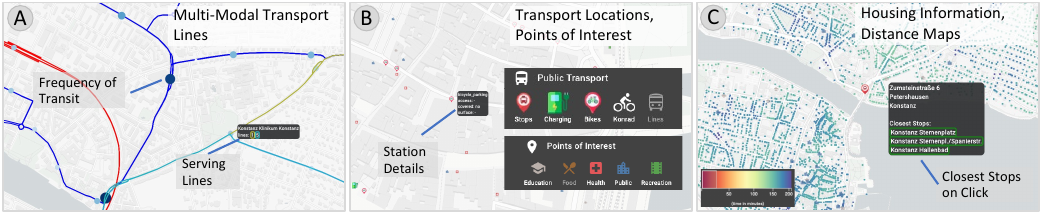}
    \caption{Data Exploration Mode: Map layers to inspect underlying information like transport lines, stop locations, points of interest, and housing data.}
    \label{fig:standard_map_layers}
    \vspace{-0.5cm}
\end{figure*}

For geographical information, we leverage \textit{OpenStreetMap (OSM) dumps} (\texttt{*.osm.pbf}) available from Geofabrik~\cite{Geofabrik.OSM.2024}, providing streets, \textbf{map tiles} for the visualization, and \textbf{VGI-POI information}.
While in theory, housing location data could be extracted from OSM, due to quality issues, we consider dedicated datasets like \textit{HK-DE}~\cite{advonlineWorkingCommittee.2024} for \textbf{housing coordinates}.
General population data for individual counties (detailing \textbf{age distribution} and \textbf{population density}), to enable demographic weighting, is retrieved from \textit{census data}~\cite{DeStatis.Zensus2022.2024}.
Figure~\ref{fig:standard_map_layers} shows the map layers that can be used to inspect the underlying source data.

\paragraph{Network Simulation and Data Pipeline}
We integrate these data sources in an analysis architecture, shown in Figure~\ref{fig:teaser}.
The different data sources are joined in a \textit{PostgreSQL} database with \textit{PostGIS}.
Data sources can be accessed adaptively as layers for visualization and serve as input to travel-time calculations.
For travel time computations, we rely on existing routing simulation tools, using r$^5$, the \textit{Rapid Realistic Routing on Real-world and Re-imagined networks}~\cite{Pereira.R5.2021} routing engine, %via the \textit{r5py} library~\cite{Fink.R5PY.2022}
leveraging the GTFS transit feed and OSM street network data.
This approach also provides walking distance estimates.%, based on shortest-path walkable path calculations.
%Later, these could also be used for demographic-specific walking speeds, to strengthen the simulation accuracy.

To save resources, we pre-compute multiple distance metrics:
(1) The N (default: 3) \emph{closest stops} to each housing location,
(2) An \emph{index of matches} between stops and transport lines, including the per-stop schedule frequencies,
(3) A \emph{travel time matrix} between stops and relevant POIs for the available modes of transport.
The pre-computed distances serve as a foundation for different group- and mode-based transport simulations, as we will outline in the following section.

\paragraph{Dynamic Network Modification}
A particular benefit of our approach is the ability to dynamically modify parts of the transport network (e.g., by placing new stops or POIs), which relies on a targeted and contained update of the network simulation.
This information is provided interactively by the user by, for example, changing schedules or moving stop positions (with semi-automatic updates to travel times and schedules), updating the directly affected areas.
However, such progressive updates might lead to compounding inaccuracies over time, in particular for indirect effects (see Section~\ref{sec:limitations_future_work}).
To solve this issue, we recompute the entire simulation in the background or on demand, e.g., between user sessions.
For more details on the dynamic modification, we refer to Section~\ref{subsec:adapt_sim}.

\subsection{Profile-Based Computation}
\label{sec:profile-based-compuations}

\begin{figure}[tbh]
    \centering
    \includegraphics[width=\linewidth, alt={The \textbf{Scenario Editor} to modify the mobility profiles of different demographic groups (B).
    For each demographic, different categories of interest (A, i.e., one or more different POI types) are modeled with a probability of visit (C) for a particular time of day within a prototypical week. The inset (D) shows a more detailed example. }]{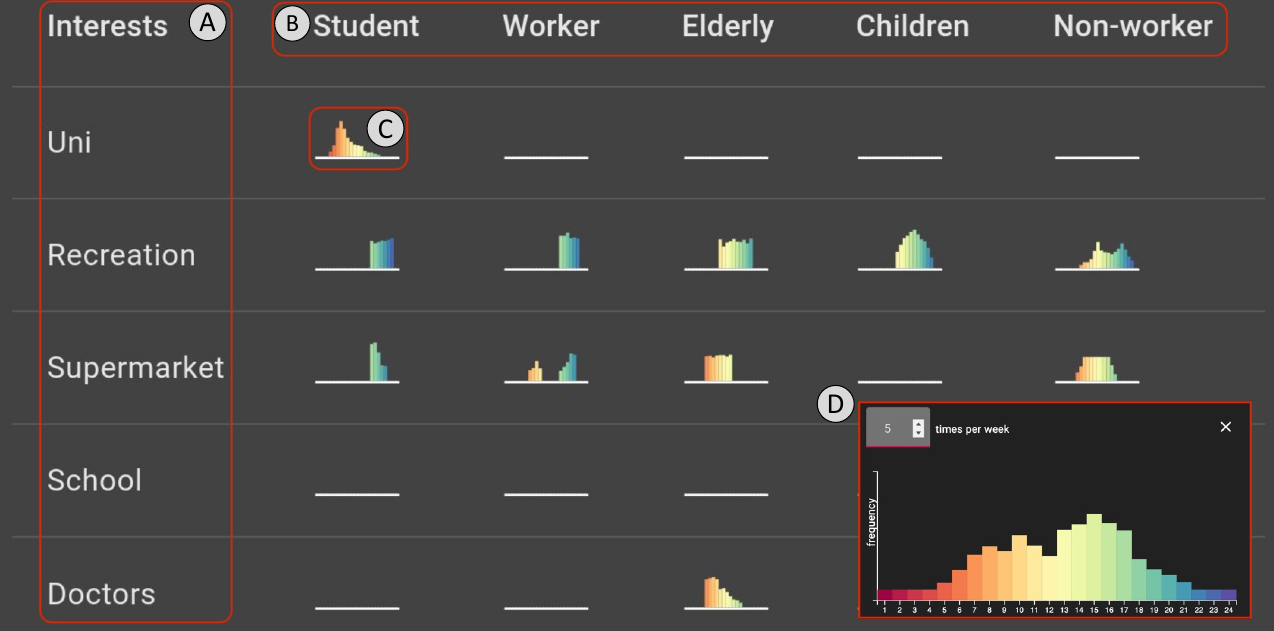}
    \caption{The \textbf{Scenario Editor} to modify the mobility profiles of different demographic groups (B).
    For each demographic, different categories of interest (A, i.e., one or more different POI types) are modeled with a probability of visit (C) for a particular time of day within a prototypical week. The inset (D) shows a more detailed example. 
    }
    \label{fig:scenario_editor}
\end{figure}

\paragraph{Mobility Profiles}
Mobility patterns and usage differ between demographics and need to be accurately modeled to achieve reliable results.
An independent modeling of each demographic allows for more user-specific results, strengthening self-identification among recipients while more accurately reflecting local demographic distributions.
To create a simulation that is easily adaptable and allows for progressive live updates, we utilize a combination of \textbf{fine-grained travel-time analysis at household level}, paired with an analysis based on group mobility profiles.
This allows us to calculate the \textbf{total travel time} for a specific demographic with their particular interests and compare these for each house, for example, through the display of this travel time or the use of weighted scores.
We decided against the use of a full agent-based simulation (i.e., simulating individual passengers or traffic vehicles on their way), which requires extensive computation. Further, such an approach would be more useful when working to create a network from scratch or considerably change it, but it is less useful when working with a primarily fixed network with only a few changes, as in our case.

Population groups and their associated profiles can be created arbitrarily by the users, for example, groups such as schoolchildren, students, working people, and the elderly.
These mobility profiles specify daily schedules within a prototypical week and frequented points of interest for these different person groups.
\begin{wrapfigure}{r}{0.15\textwidth}
\vspace*{-0.75cm}
  \begin{center}%
    \includegraphics[width=\linewidth, alt={A more detailed Mobility Profile card example for a group \textit{Elderly}}]{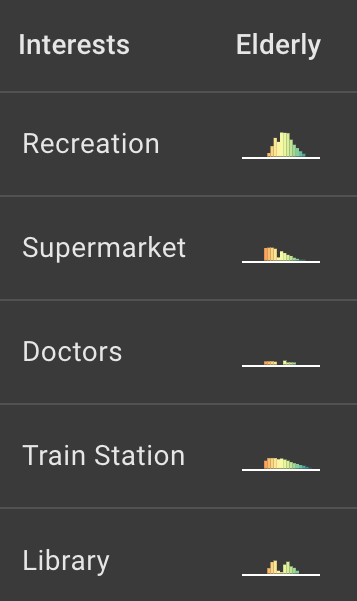}%
  \end{center}%
  \vspace*{-4mm}
  \caption{A more detailed Mobility Profile card example for a group \textit{Elderly}}
  \label{fig:mobility_profile_card}
  \vspace*{-0.4cm}
\end{wrapfigure}
For the default profiles, we use statistical data obtained from large population surveys like Mobility in Germany~\cite{MiD.2018}, which details daily activity patterns for such groups, but this is adaptable when detailed statistics or estimates are available.

Figure~\ref{fig:scenario_editor} shows the \textit{Scenario Editor}, in which users can interactively modify the target locations, in particular POIs, that these different demographic groups visit, and the usage times that these groups use public transport to get these locations.
In the profile editor, demographic groups can be added or refined; for example, add a new category of \textit{tourist} or split existing groups.
These points of interest are chosen from pre-defined groups (such as recreational areas, supermarkets, or doctors) or single locations (e.g., universities).

For each POI group, we can specify the frequency with which we expect the demographic group to travel to/from the points of interest, with one likelihood value for each hour of the day.
For example, we can assume that children collectively travel to school in the morning (between 7:00 and 9:00) and then leave between midday or late afternoon.
Students travel to/from the university more spread over the day and use recreational POIs and bars in the evening.

\paragraph{Route Selection}
To compute a transport coverage score, we compute the distances to a set of POIs based on the demographic mobility profiles.
However, depending on the type of POI, it is fair to assume that passengers exhibit different behaviors:
For some categories, passengers might only frequent a specific location, e.g., the university, whereas, for others, passengers might want to visit multiple different locations.
Besides the frequency and times, we can define three different types of POI sampling:
\begin{itemize}
    \item \textbf{Near Location Sampling}: For some types, like shops or supermarkets, it's reasonable to assume that passengers are more likely to choose a nearer location as their target, for example, the nearest supermarket or the nearest organic supermarket of a chain.
    \item \textbf{Random Location Sampling}: Passengers travel to a random choice within a POI group within their home area, e.g., a reasonable assumption for recreational areas or higher-end restaurants.
    \item \textbf{Specific Location Sampling}: A user might frequent a single, specific location (like the University or the main train station), in particular, if the location type is unique. \emph{Note:} This is a special case of the random location with a group size of one.
\end{itemize}
Note that some POI groups can make sense for either Near Location or Random Location sampling, e.g., the \textit{Italian restaurant around the corner} vs. \textit{the upper-class restaurants in town} for business meetings.
When computing scores for each household, these choices are considered in sampling for the route selection and travel time calculation and contribute to the overall coverage score of a location based on the generated mobility profile. 

\paragraph{Travel Time Calculation}
These profiles are used to generate individual transport coverage estimations, i.e., to compute how well different residence locations are covered by different modes of transport and how well they are suited for different demographic groups. To quantify the coverage, we aggregate the travel times for each location and demographic group, i.e., a measure that summarizes how much travel time a person spends on travel to/from their residence to relevant points of interest over a prototypical week.
To simplify the computation, we use the pre-computed \textit{travel time matrix}, i.e., estimated travel distances between all stops and points of interest) as the basis. 
For each demographic group, we then compute the estimated average weekly travel times by considering the mobility profiles and considering how often the person visits a location per week. We selected frequency $I$ (one hour), with the entries of the travel time matrix $T$ to get the expected average daily travel time for a connection between residence and point of interest:
$$
    \mathbb{E}[t_{\textrm{poi},\textrm{day}}(P, stop)] = \frac{1}{|I|} \sum_{h \in I} f_{h}(P) * T_{\textrm{stop},\textrm{poi}}
$$
With $P$ being the group profile, $f_h$ being the frequency of travel at a certain time of day $h$, and $T_{\textrm{stop},\textrm{poi}}$ being the entry of the travel time matrix for a given time of the day, i.e. the travel time between a stop and a point of interest based on the $r5$-simulation. For a given demographic group, we then compute a weighted average based on the times per week we expect a person to frequent points of interest:
$$
    t_{\textrm{total}, \textrm{week}}(P,stop) = \frac{1}{\#\text{visits}_{total}} \sum_{poi \in POIs} \#\text{visits}_{week} * t_{\textrm{stop}}(P,poi)
$$
$t_{total,week}(P,stop)$ thus indicates the estimated average travel time a person spends in a typical week when commuting from a given stop. To compute the total weekly travel time, we add the computed walking times between a residence and stops for each trip.

We end up with a simple, single travel time estimate for each location and demographic group. We can further refine our score to measure the quality of existing infrastructure by considering population/census data, which allows us to estimate the distribution of demographic groups within different neighborhoods.

\subsection{Adaptable Simulation}
\label{subsec:adapt_sim}
One benefit of our approach is the ability to update the simulation, which enables a flexible analysis of alternative scenarios and changes to the traffic infrastructure. We allow for three types of edit: (1) \textit{Add:} Users can place new points of interest, housing, and stops to street locations, i.e., directly next to an existing street, (2) \textit{Move:} Users can move around stops potentially influencing lines, (3) \textit{Remove}: Users can delete locations. We want to note that at this point, certain special cases, like completely removing a bus line, are not fully covered, and we only assume small modifications to the existing infrastructure, like placing an additional stop or point of interest.

To enable this type of dynamic manipulation, a map/data layer is cloned when a human starts editing. Editing is done on this cloned layer, and thus, the state of the simulation can always be reversed to the original state. When placing/moving or removing a POI of the simulation, a re-computation of directly affected travel times is triggered. We trigger these re-computations in an area bound by the Voronoi area surrounding the POI or stop, i.e., we select all residences closest to the edited POI or stop, compared to other POIs or stops of the same category and compute the total travel times. Thus, the changes are only localized and can often be done relatively quickly. At the current state, the adapted travel times for a bus when placing a new station on a line are semi-automated, where users have to manually confirm updated scheduled times. Considering potentially resulting rippling effects and completely automating (using the same routing backend), updating the schedule accordingly would be possible in a more complex agent-based simulation in the future.

\section{Visual Analytics Application}

\label{sec:system}
In this section, we present the design of a visual analytics application that enables interaction between the human, data, and simulation, enabling multiple analysis modes.
In the previous section, we have described the underlying simulation system, which informs three specific requirements for our agent-based public transport analytics system: (R1) Visualizing the underlying data and geospatial information informing the travel time simulation, (R2) the automated generation of a simulation based on publicly available data, and (R3) the interactive adaptation of the simulation, e.g., by placing new stops and modifying existing locations.
We start with an overview of the main user interface before addressing how our design meets these requirements.

\begin{figure*}[tbh]
    \centering
    \includegraphics[width=\linewidth, alt={The application allows a comparison of travel times for different demographics, based on generated mobility profiles. We can identify neighborhoods that are especially suited or unsuited for different demographic groups. This indicated the potential for targeted modifications.}]{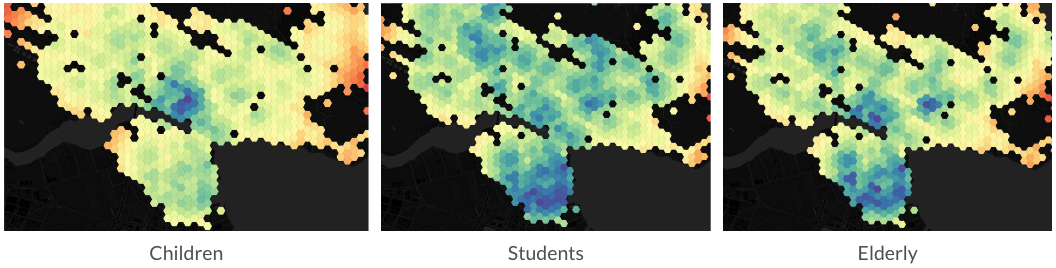}
    \caption{The application allows a comparison of travel times for different demographics, based on generated mobility profiles. We can identify neighborhoods that are especially suited or unsuited for different demographic groups. This indicated the potential for targeted modifications.}
    \label{fig:demographic_comparison}
\end{figure*}

\subsection{User Interface}
The main view of the application is shown in Figure~\ref{fig:user_interface} and consists of two main components: a map view with a legend (a) and a sidebar (b-f) with several control components.

\textbf{Main Map View} --- %
When starting the application, the user first selects a region, such as a county or state, on which the analysis is focused.
Then, a map view of the selected region is loaded (using openly available tile sets), where the map view serves as the context for multiple selectable overlays and analysis layers to show or hide information such as public transport infrastructure, points of interest, roads, or heatmap layers.
A legend adapts to the information shown and provides relevant context and color/normalization options.
The user starts their exploration in the default mode: the (1) \textit{Data Exploration Mode}. Via the sidebar, they can also navigate to the (2) \textit{Simulation Analysis Mode}, and (3) \textit{Editing Mode}.

\textbf{Side Panel and Layers} --- %
The overlay controls (b) in the sidebar offer map controls, public transport layers, background information (like houses, population densities, or streets), and different POI categories that can flexibly be blended into the map view.
The analysis scenarios (d) reflecting the mobility of a set of demographics can be modified through a separate editor (see below) as well as compared to different assumptions (c), and some of the aspects are interactively exposed for quick analysis (e). The menu dynamically adapts to the available information, depending on the data source.
Further general settings are also available (f).

\textbf{Scenario Editor} --- %
Editing the scenarios of the individual demographics is possible by using the Scenario Editor, shown in Figure~\ref{fig:scenario_editor}.
It displays a range of customizable demographics and at which times they are likely to use public transport for a range of interests, coupled with a set of POIs.
Both dimensions are editable, and the usage likelihood can be freely chosen for the specific simulation intervals within a prototypical week.
For a detailed description of the underlying concepts and possibilities, we
refer to our description in Section~\ref{sec:profile-based-compuations}.

\subsection{Data Exploration Mode}
The goal of the \textit{data exploration mode} is the analysis of the existing public transport infrastructure, as well as the underlying data that feeds into the transport simulation.
The information available in the data exploration mode is shown in Figure~\ref{fig:standard_map_layers}.

To accommodate our target group of lay users, we chose a simple layer-based interface with information that can be toggled on/off on demand via a legend. Multiple map styles and sequential color scales can be selected in the application to accommodate diverse user preferences and color vision deficiencies. Information that can be displayed is grouped into several categories: \textit{Public transport infrastructure} (ranging from bus stops to cycle parking spots, BEV charging stations), \textit{background information} (route networks, housing, population data or bus stop frequencies), \textit{Points of interest} (schools, medical facilities, recreational areas, etc.). Potential other data sources, like bike or car-sharing stations, can be integrated and shown.

\subsection{Simulation Analysis Mode}
\label{subsec:sim_analysis_mode}
The \textit{Simulation Analysis Mode} enables the analysis of the transport simulation results.
The simulation enables the analysis of transport coverage for different modes of transport, geographical locations, and demographic groups.

The results of the simulation are provided as additional color-heatmap layers displayed on top of the exploration map. The output of the simulation is travel time estimates for locations, which are mapped to color values on a sequential color scale. The travel time estimates are dependent on the underlying data and mobility profiles. The results can be displayed at different levels of granularity based on the zoom level: At low zoom levels, the results are aggregated via averaging of travel times and displayed in a hexagonal grid. The size of the hexagonal tiles changes dynamically based on the zoom level. At high zoom, results for single residences are visible on the map, and no aggregation takes place. The color scale is dynamically adjusted between the maximum and minimum values in the visible map area. This allows users to identify more fine-grained differences at higher zoom levels. The color scale legend is always visible to indicate these changed scales.

The map menu allows users to choose which simulation results to display, as multiple filtering and aggregation options are available. It is possible to choose between travel time estimates per \textit{demographic group} based on the selected mobility profiles, \textit{modes of transport}, selection of \textit{point of interest}, i.e. a (set of) potential destination(s):
\begin{itemize}
    \item An aggregated travel time estimate for each location that is based on a weighted average of travel time estimates for all demographic groups, based on population proportions, with the full set of \textit{POI}s and modes of transport being considered
    \item Travel Time-heatmap per demographic group: By selecting only a specific group, we can visualize individual heat-maps for each group to identify areas that are more/less suited for particular demographics
    \item Travel Time estimates for specific \textit{POI}s: We can analyze the travel times towards particular destinations (e.g., the university or main train station of a city), to evaluate how well specific neighborhoods are connected to specific POI
\end{itemize}
Because all of these travel estimates are computed during the creation of the full network simulation, they are readily available when filtering the simulation result heatmaps. In very large areas, this can come with considerable storage requirements.

\subsection{Editing Mode}
\label{subsec:editing_mode}
Finally, the \textit{editing mode} enables the dynamic and interactive adaptation of the data, such as placing new stops, points of interest, or homes.
Furthermore, we can edit the mobility profiles of demographic groups to steer the group-based analysis. 

\begin{figure*}[tbh]
    \centering
    \includegraphics[width=\linewidth, alt={Our system enables users to interactively place transport infrastructure, \textit{POI}s, and housing to explore the effects of new or updated infrastructure. In this scenario, a new bus stop is placed in a previously less connected neighborhood. In this scenario only one bus line is affected which limits the effect on other locations. Together with changes in the mobility profiles, our tool enables fast and accessible public transport planning.}]{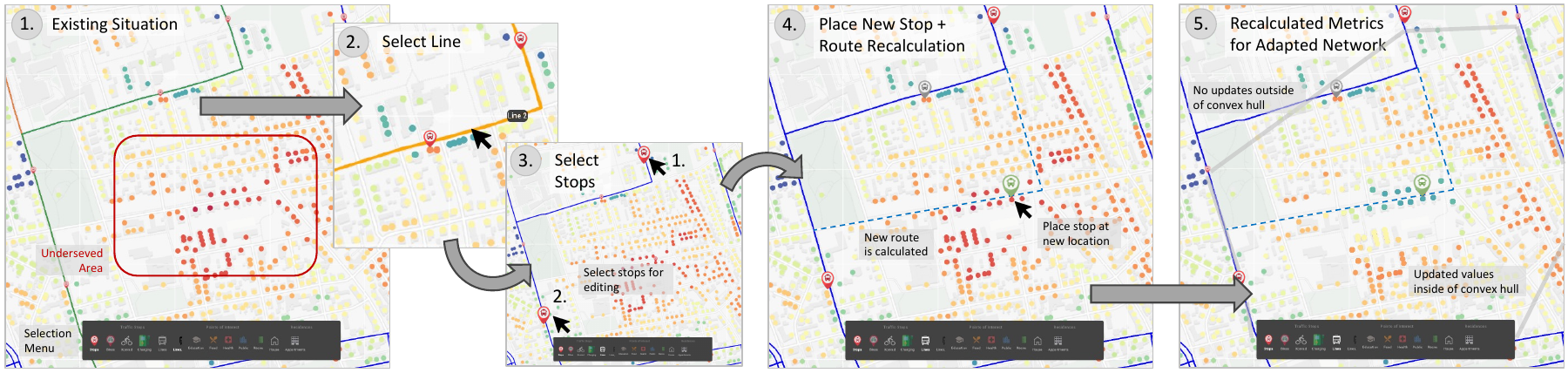}
    \caption{Our system enables users to interactively place transport infrastructure, \textit{POI}s, and housing to explore the effects of new or updated infrastructure. In this scenario, a new bus stop is placed in a previously less connected neighborhood. In this scenario only one bus line is affected which limits the effect on other locations. Together with changes in the mobility profiles, our tool enables fast and accessible public transport planning.}
    \label{fig:process_time_recomputation}
\end{figure*}

As already outlined in Subsection~\ref{subsec:adapt_sim}, there exist three ways to adapt the geographic data underlying the simulation that need to be supported by user interactions: Adding, moving, or deleting elements.\\
When entering the editing mode, a new menu bar with all placeable elements appears at the bottom of the view. By clicking on an element, e.g., a POI, bike sharing station, etc., users can place new elements on the map. When placing a new stop for a transport route, the transport line network overlay is activated; users first select a transport line (e.g., a bus line) and then place a stop on the map. The bus route is then rerouted visually to the newly placed stop. This process is shown in Figure~\ref{fig:process_time_recomputation}. As we have noted before, in the current system, the placement does not yet affect travel times and schedules, i.e., we assume that stops do not affect travel times. A more sophisticated dynamic stop placement and route simulation is planned as a major future improvement.
Similarly, by drag and drop, the position of map elements can be modified. Finally, users can activate a \textit{deletion mode}, in which double-clicking of map elements removes those from the map. All of these changes trigger limited recomputation of the simulation as described above.

The second notable part of the editing mode is the mobility profile editor (see Figure~\ref{fig:scenario_editor}). The editor is contained in a separate modal.
The user can view a grid of demographic groups and \textit{POI} classes. The editor also allows access to defining new groups (including values like individualized walking speed), as well as editing of \textit{POI} categories (i.e., which POI from data sources such as Open Street Map are grouped). For each combination of demographic group and POI category, a user can edit the amount of times a person visits a POI weekly and with which frequency per hour of day. We call these assignments \textit{mobility profiles} for particular groups. Since changes to the mobility profiles are not localized geographically, they require a more complete re-computation of travel times. Therefore, we delay computation until the user confirms all pending changes in the editor.

\section{Case Study}
\label{sec:case_study}
Our system is aimed primarily at policymakers and non-specialists who have an interest in the analysis and improvement of public transport infrastructure but do not have access to more detailed closed data and do not have the resources to work with numerical flow simulation planning tools.
Such users might include political activists, lobbying groups, journalists, or interested citizen scientists.
We argue that a broader representation of interested and supporting tools can strengthen the quality of public discourse and decision-making.

In the following, we present two case studies that demonstrate our developed application for public infrastructure analysis and planning.

\subsection{Case 1: Identifying Underserved Areas/Demographics}
\textbf{(G1, G2)} As a first case study, we look at the \textbf{county of Konstanz, Germany}, an area with c. 280.000 inhabitants.
We outline the steps taken to identify neighborhoods of the city that are not covered well by existing infrastructure, i.e., have high travel times to relevant destinations.
In particular, we want to look at the situation for different demographic groups.
The outcome can be heatmaps indicating travel times, which can be passed to policymakers or administrators as a way to start a discourse.

Emilia is part of a local environmental advocacy group and wants to perform a data-driven investigation into insufficient public transport access, initiated by complaints she received.
She selects the region of the region of "Konstanz" in the tool and is greeted by the main map-based interface (Figure~\ref{fig:user_interface}).
She first zooms into the area of the city that interests her.
By clicking on the layer icons for infrastructure and POIs (Figure~\ref{fig:user_interface}\textbf{B}), she inspects the existing data and confirms that the density of transport infrastructure in a certain area seems relatively low. She activates the walking distance heatmap (Figure~\ref{fig:standard_map_layers}\textbf{C}), which shows for each individual location how far the distance to the next bus stop or tram station is.
She notices that there is a set of streets that exhibit a relatively high walking distance, and the next stop is only infrequently served.

She has the suspicion that this might, in particular, affect elderly people in their daily routine:
She switches to the \textit{simulation analysis mode} (Subsection~\ref{subsec:sim_analysis_mode}) and reads through the default demographic profile for her group (Figure~\ref{fig:mobility_profile_card}), which is based on official federal survey data on mobility preferences \cite{MiD.2018}.
She is satisfied with the existing profile, performs the travel time simulation, and switches the main view to the demographic heatmap.
Here, she toggles between the aggregated heatmap and the group-specific maps (Figure~\ref{fig:demographic_comparison}).
She notices that the area of interest indeed seems less well connected overall compared to much of the surrounding neighborhoods.
%Travel times are on the higher end in similarly located neighborhoods.
When switching to the heatmap, she notices that travel times for elderly are especially high there, confirming her suspicion.
She especially identifies a set of housing blocks that are especially disconnected.

Emilia exports these heatmaps as images, together with a list of streets and approximate travel times, as well as a view of the utilized mobility profiles to back up her claims.
She prepares to raise this issue with the local council and transport company, starting the discussion on improving access for the elderly and pushing for change.

\subsection{Case 2: Evaluating the Effect of Infrastructure Changes}
\textbf{(G2, G3)} But how could such change look like, and how can one use our approach towards modeling and recommending potential changes to the transport infrastructure?
For this case study, we chose the smaller city of Konstanz, Germany, with a population of around 85,000 inhabitants.

\begin{figure}[tbh]
    \centering
    \includegraphics[width=\linewidth, alt={A difference map visualization showing additional types of analysis possible with the mobility profile editor: (a) the comparison of connectivity at night-time compared to day-time (white: no difference, magenta: large difference) (b) advantage of car usage over reliance on public transport (white: no difference, green: large difference).}]{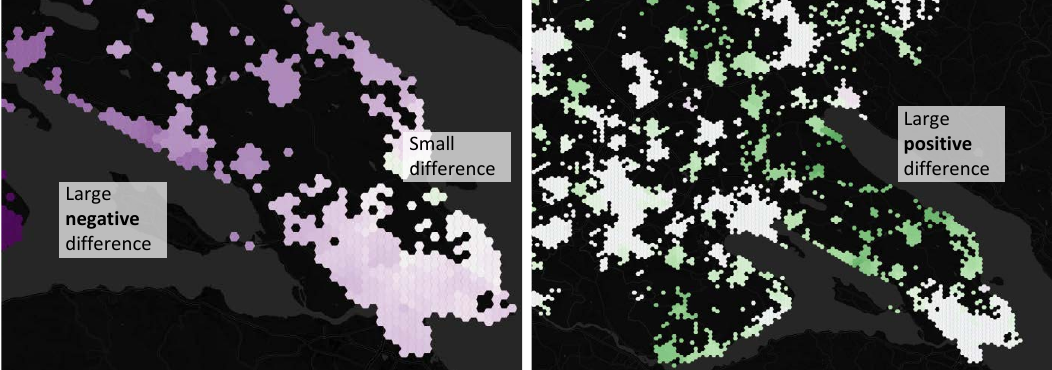}
    \caption{A difference map visualization showing additional types of analysis possible with the mobility profile editor: (a) the comparison of connectivity at night-time compared to day-time (white: no difference, magenta: large difference) (b) advantage of car usage over reliance on public transport (white: no difference, green: large difference).}
    \label{fig:difference_map}
\end{figure}

Alex is a student at the local university and is often frustrated that his home takes a long time to reach.
He wonders if there are possible improvement opportunities in his neighborhood.
Similar to the previous case, Alex performs a heatmap analysis, where he notices that bus lines stop on the very edge side of his neighborhood and hypothesizes that rerouting a bus route through its center might improve travel times for many houses.
After identifying possible locations, he switches to the \textit{editing mode} tab (Subsection~\ref{subsec:editing_mode}).
In the editing mode, the user can choose to place elements such as POIs, bike sharing, or new transport infrastructure (Figure~\ref{fig:process_time_recomputation}).
Alex wants to evaluate the effect of a changed bus stop placed at a strategic location.
He first selects the bus stop icon in the menu bar.
The bus line overlay is activated, and he can click on an existing bus line to move one stop from this bus line.
After clicking on a bus line, he moves the stop from the edge to the center (Figure~\ref{fig:process_time_recomputation}.\textbf{2/3}).
Now, the simulation is re-computed in two stages, indicated by a loading icon: Within a region constrained by the neighboring bus stop locations, walking times to the closest location are re-computed to consider the change.
Simultaneously, the travel times from the new bus stop to the relevant POIs are recomputed (i.e., a new row to the travel-time matrix is computed), as described above.
Finally, for all relevant points, the travel time estimates are updated and are displayed on the travel time map (Figure~\ref{fig:process_time_recomputation}.\textbf{5}).
To evaluate the effect of the change, the user can select a difference visualization, i.e., a heatmap that displays the differences in travel time estimates with and without the modification (compare to Figure~\ref{fig:difference_map}).
Alex tries out multiple placements in order to identify the best possible location.

After a few tries, Alex is satisfied with the chosen location.
He decided to present the analysis case on his own personal blog, supported by the generated visualizations.
The blog gets picked up by a local newspaper and starts a discussion about the local public transport.

\subsection{Initial User Study}
\label{sec:user_study}
We conducted a initial user study with eleven participants recruited from a university environment, aged between 21 to 32 years: six computer science (CS) PhD students (PHD1-6), three CS master students (CS1-3), one finished CS master (CS4), and one history student (HS1).
Of those, ten had prior knowledge of visual analytics, and seven had previously worked with GIS tools.
The study was conducted in a university lab in a controlled environment, with a short preliminary questionnaire, a short introduction, a task-based assessment using think-aloud, followed by NASA TLX and User Experience Questionnaire (UEQ) as well as a semi-structured interview at the end, taking in total about one hour.
The setup was shown on a 14.2-inch display with a resolution of 3024x1964, using the web-based prototype.

The tasks were (1) initial exploration and finding specific layer information, (2) creation of a new scenario with profiles and interests, (3) exploring the heatmap to identify areas with diverging transport coverage, and (4) creating a second scenario and comparing the two.

Participants successfully completed these tasks, and the NASA Task Load Index (TLX) remained low, except for a medium demand on mental capacity.
The UEQ scores remained high, in the range $+1.5$ to $+2$ for all categories, with the exception of slow/fast ($-0.3$).
Suggested improvements focus on the scenario editor in the way demographics, interests, and times are edited.
In particular, the use of more detailed agent profiles for better accuracy was suggested, from which to start modification to ease the burden on the user.
The overall feedback was very encouraging, and the users highlighted the prototype's usefulness, ease of exploration, and available information through different layers to explore in detail their specific scenarios, but seeing the potential for further improvements in filtering for specific questions.
The main heatmap view was well regarded, often matching the user's suspicions or experiences, only suggesting further options like more transparency options and more information on demand, like population data in tool tips and visual routes for default agents.

\section{Discussion and Future Work}
\label{sec:discussion}
The case studies have shown the comparative ease with which a scenario-based analysis of existing public transport infrastructure can be generated, the effects of change can be analyzed, and the insights that can be drawn from it, which is validated by the initial user study.
In the following, we want to highlight some key findings before discussing the limitations of our approach.

\subsection{Findings and Lessons Learned}
\label{sec:findings}
We outline how our system matches the three goals of (G1) visualizing the quality of existing infrastructure, (G2) enabling detailed analysis for different demographic groups and usage scenarios, (G3) simulate the approximate effectiveness of modifications in the case studies (see Section~\ref{sec:case_study} and initial user study (see Section~\ref{sec:user_study}).
This is achieved via simple yet expressive visualizations (R1), a simulation based on open-source real-world data (R2), which is easily adaptable (R3).

For this, we found that \textbf{detailed, default agent profiles} are important to foster user trust and understanding, as their manual re-creation is labor-intensive and primarily important for specific investigations.

Overall, we demonstrated that a \textbf{data-driven approach}--only relying on open data--and using aggregated travel-time simulations through mobility profiles can already lead a long way to generate \textbf{actionable insights}, revealing quality differences between neighborhoods in a comparable way, in particular studying demographic effects, which indeed match with actual user experiences.
Through this, we are able to visually identify (inherent) weak points in a city's infrastructure and support users in exploring alternatives.

\subsection{Limitations and Future Work}
\label{sec:limitations_future_work}
A limitation of our current approach is the limited accuracy of how daily routines and POI visits \textbf{accurately reflect the mobility patterns} of the population.
The default provides a relatively coarse number of demographic groups and POI-based interests, which can, however, be refined by the user with more detailed activity data.
While these movement patterns (both in the default and manually refined detailed version) are based on actual large-scale surveys~\cite{MiD.2018}, particularities and special cases like going for a walk, spontaneous trips, or a travel choice influenced by external factors might lead to noticeable divergences.
By design, such a travel-time-based analysis can only give imperfect simulation results.
This could be overcome with more detailed data available, for example, through volunteered GPS-based tracking or more-detailed phone tracking through cellular providers or platform providers like Google or Apple, which, however, raises privacy concerns.

Our approach currently only includes \textbf{target timetable data}, not (commonly occurring) delays, canceled trips, or real-time information that diverges from the schedule.
This allows our approach to find \textit{inherent structural issues} in the network but makes it harder to detect issues stemming from external factors, like regular traffic jams, when schedules are unrealistic and do not factor in these aspects already.
Therefore, our approach strongly depends on the accuracy of the timetables throughout the day.
In the future, we could use retrospective actual data (when available) or use the editing mode to simulate such delays.

The \textbf{modifications of the transport network} does not consider secondary effects in our existing simulation framework.
Right now, for example, when adding a new stop or changing the schedule, it would make sense to perform changes to timetables also of \textit{other} lines, adapting their schedules, too, to ensure smooth connections and reduce waiting times.
Progressive schedule changes are currently not fully supported because they can quickly explode and would necessitate complex adaptation of timetables and routes, requiring an almost complete (and slow) re-simulation of the whole area, which is not compatible with the interactive approach.
Our system could be extended to integrate other agent-based simulation frameworks like \textit{MatSim}~\cite{Axhausen.MATSim.2016} or \textit{Sumo}~\cite{DLR.SUMO.2024} or utilize more computationally extensive agent-based simulations.

Another promising avenue to improve the editing system would be to introduce recommendations for possible modifications, i.e., the identification of the best possible locations for the placement of new infrastructure and/or POIs. This poses an interesting optimization problem due to a potentially large search space.
As a possible intermediate step before the fully automated generation of recommendations, one could explore the evaluation in multiple locations close to a selected edit in order to speed up the comparison of multiple possible modifications.

Our system is designed to be accessible and user-friendly, especially for non-expert users. While we expect some time required to become familiar with all functionalities, we have already \textbf{evaluated} the usability for non-expert users in a initial user study with eleven participants, indicating its usability for individuals or interest groups. The usability of the approach could be investigated in a quantitative user study. This could open up public transport planning to wider audiences and improve the quality of discussion for integrating civil society into planning processes. We can imagine our approach to enable even \textbf{broader participation of citizens}, e.g., we could use the mobility profile editor to collect survey data from citizens and aggregate the results to get more accurate profiles in areas where surveys have been limited so far.

The interactive nature requires some \textbf{computing capability} for a user, which is achievable for a domain expert setting but becomes a scalability issue when making the system open to a general audience.
In the future, one could imagine a public \textit{read-only version} of the system, while interactivity would be limited to a restricted user group, for example, through prior registration and application.
Due to similar reasons, \textbf{multi-modal and cross-modal transport} is not yet considered, for example, combining public transport with individual mobility like a car.
Our approach so far \emph{only} considers public transport in itself, which might not always reflect actual usage (e.g., taking the bike to the train station), but this could be added with more compute available.
When used by city planners and other expert users, integration with de-facto standard frameworks like \textbf{ArcGIS} might also be desirable.

Right now, we have used the application to model traffic within Germany.
While some sources (like OSM) have \textbf{international coverage} and GTFS is a worldwide standard for network schedules, we expect the need to adapt some data to other regions, for example, census data. However, the use of standards should simplify these adaptations.

\section{Conclusion}
\label{sec:conclusion}
We have introduced a novel agent-based simulation approach for analyzing and optimizing public transport infrastructure at housing-level granularity.
By leveraging housing locations, transport networks, and schedules combined with agent-based simulations, controlled through rapid iterative refinements, our system provides detailed insights into the accessibility and connectivity of public transport networks for members of the public.
This approach not only bridges the gap between theoretical planning and practical implementation but also identifies inherent structural issues, thereby highlighting specific areas where interventions are most needed, thus facilitating more informed urban planning decisions.
Our findings through a case study and a initial user study, indicate that our approach can effectively identify critical gaps in public transport services and suggest enhancements that are both practical, focusing on economically viable improvements rather than comprehensive overhauls.
By enabling detailed and scalable analyses, our system assists policymakers and the interested public, but potentially also city planners, in making data-informed decisions that foster a more equitable and efficient mobility transition.
Looking forward, we see substantial opportunities for extending our approach to include (inter-) national data, which could further refine the accuracy and timeliness of transport planning across regional boundaries.
By providing both comprehensive insights and support the generation of actionable paths to improvement, our contribution paves the way for more strategic, data-driven urban transport planning and systems, adapting to changing urban landscapes and mobility demands, ultimately contributing to more sustainable cities and improved public infrastructure.

\acknowledgments{%
This work has been received funding by the Deutsche Forschungsgemeinschaft (DFG, German Research Foundation) under Germany's Excellence Strategy – EXC 2117 – 422037984.
}

\bibliographystyle{abbrv-doi-hyperref}

\bibliography{references}

\end{document}